\documentclass[]{aa}

\usepackage {graphicx}
\usepackage {times}

\def\apj{\,{\rm ApJ}}
\def\apjl{\,{\rm ApJL}}
\def\apjs{\,{\rm ApJS}}

\def\aap{\,{\rm A\&A}}

\def\mnras{\,{\rm MNRAS}}

\begin{document}


\titlerunning{Are E+A Galaxies Dusty-Starbursts?}
\authorrunning{T. Goto}

\title{Are E+A Galaxies Dusty-Starbursts?: VLA 20cm Radio Continuum Observation}
\author{Tomotsugu Goto }
 \institute{Department of Physics and Astronomy, The Johns Hopkins
  University, 3400 North Charles Street, Baltimore, MD 21218-2686, USA\\
              \email{tomo@jhu.edu}
             }


\label{firstpage}

   \date{Received May 07, 2004; accepted }

   \abstract{
 E+A galaxies are characterized as a galaxy with strong Balmer
 absorption lines but without any [OII] or H$\alpha$ emission
 lines. The existence of strong Balmer absorption lines means that E+A
 galaxies have experienced a starburst within the last $<$1-1.5 Gyr. However, the
 lack of [OII] or H$\alpha$ emission lines indicates that E+A galaxies
 do not have any on-going star formation. Therefore, E+A galaxies are
 interpreted as a post-starburst galaxy. For more than 20 years,
 however, it has been a mystery why E+A galaxies underwent a starburst then stopped abruptly.

 One possible explanation to E+A galaxies is the dusty starburst scenario, where E+A galaxies have on-going star formation, but optical emission lines are invisible due to the heavy obscuration by dust.
To test this dusty starburst scenario, we have observed 36 E+A galaxies carefully selected from the Sloan Digital Sky Survey Data Release 1 in 20cm radio continuum using the VLA. Since the radio emission is not affected by the dust extinction, the star formation rate (SFR) in dusty galaxies can be revealed by the 20 cm continuum. All of our 36 target E+A galaxies are selected to  have H$\delta$ equivalent width greater than 6\AA\ and no detection of [OII] or H$\alpha$ emission lines within 1 $\sigma$. These selection criteria are much stronger than previous E+A selections in our attempt to select galaxies in the pure post-starburst phase without any remaining star formation.
 Except for the two galaxies with a nearby radio source, none of our 34 E+A galaxies are detected in 20 cm continuum to the limits reported in Table 1. 
 The obtained upper limits on the radio estimated SFR suggest that E+A galaxies do not possess strong starburst ($>$ 100 M$_{\odot}$ yr$^{-1}$) hidden by dust extinction for the whole sample while 15 ($z<0.08$) E+As have lower SFR upper limits of $\sim$ 15 M$_{\odot}$ yr$^{-1}$.

   }


   \maketitle



\section{Introduction}\label{intro}

  Dressler \& Gunn (1983; 1992) discovered galaxies with mysterious spectra while
 investigating high redshift galaxy clusters.
 The galaxies had strong Balmer absorption lines with no
 emission in [OII]. These galaxies are named 
 ``E+A'' galaxies since their spectra looked like a superposition of that of
 elliptical galaxies (Mg$_{5175}$, Fe$_{5270}$ and Ca$_{3934,3468}$
 absorption lines) and that of A-type stars (strong
 Balmer absorption)\footnote{Since some of E+A galaxies are
 found to have disk-like morphology (Franx 1993; Couch et al. 1994;
 Dressler et al. 1994; Caldwell et al. 1997; Dressler et al. 1999),
 these galaxies are sometimes called 
 ``K+A'' galaxies.  However, Goto (2003) found that their E+A sample with
 higher completeness has early-type morphology. Following this discovery, we call
 them as ``E+A'' throughout this work.}.  
  Since the lifetime of A-type stars is about 1 Gyr, the existence of strong Balmer absorption lines shows 
 that these galaxies have experienced starburst within the last
 Gyr.  However, these galaxies do not show any sign of on-going star
 formation as non-detection in [OII] emission lines
 indicates.  
   Therefore, E+A galaxies are interpreted as a post-starburst galaxy,
 that is,  a galaxy which has undergone a truncated starburst (Dressler \& Gunn 1983,
 1992; Couch \& Sharples 1987; MacLaren, Ellis, \& Couch 1988; Newberry
 Boroson \& Kirshner 1990;  Fabricant, McClintock, \& Bautz 1991; Abraham et al. 1996).
  The reason why they underwent a starburst then abruptly stopped still remains one of 
  the mysteries in galaxy evolution. 
 Since starburst-driven galaxy
 transformations are an important part of galaxy formation theories
 (e.g. Somerville et al. 2001), it is important to
 understand what triggered the starburst in these galaxies and, perhaps more
 importantly, why star formation subsequently ceased so abruptly.

  At first, E+A galaxies were found in cluster regions, both in low
  redshift clusters (Franx  1993; Caldwell et al.  
  1993, 1996;  Caldwell \& Rose 1997; Castander et al. 2001; Rose et
  al. 2001) and high redshift 
  clusters (Sharples et   al. 1985; Lavery \& Henry 1986; Couch \&
  Sharples 1987; Broadhurst,   Ellis, \& Shanks 1988; Fabricant,
  McClintock, \& Bautz 1991; Belloni   et al. 1995; Barger et al. 1996; Fisher et
  al. 1998; Morris et al. 1998; Couch et al. 1998;   Dressler et al. 1999). Therefore, a
  cluster specific phenomenon was thought to be responsible for the
  violent star formation history of E+A galaxies. A ram-pressure
  stripping model (Spitzer \& Baade 1951, Gunn \& Gott 1972, Farouki \&
  Shapiro  1980; Kent 1981; Abadi, Moore \& Bower 1999; Fujita \& Nagashima 1999;
 Quilis, Moore \& Bower 2000; Fujita 2004; Fujita \& Goto 2004)
  may first accelerate star formation of cluster galaxies and later turn it
  off as well as tides from the cluster potential (e.g., Fujita 1998) and the evaporation of the cold gas (e.g., Fujita 2004).
 However, recent large surveys of the nearby universe found many E+A
  galaxies in the field regions (Goto 2003; Goto et al. 2003a, G03 hereafter; Quintero et al. 2004).  
 At the very least, it is clear that these E+A galaxies found in the
  field region cannot be explained by a physical mechanism that works
  in the cluster region. E+A galaxies have been often thought to be 
  transition objects during the cluster galaxy evolution such as the
  Butcher-Oemler effect (e.g., Goto et al. 2003b), the morphology-density
  relation (e.g., Goto et al. 2003c; Goto et al. 2004), and the correlation between various properties of galaxies with the environment (e.g. Tanaka et al. 2004). However, explaining cluster galaxy evolution using E+A galaxies may not be realistic anymore.

Alternatively, galaxy-galaxy interaction has been known to trigger
      star formation in the pair of galaxies (Schweizer 1982; Lavery \&
      Henry 1988; Liu \& Kennicutt 1995a,b; Schweizer 1996).
         Oegerle, Hill, \& Hoessel(1991)
       found a  nearby E+A galaxy with a tidal feature.   High
      resolution imaging of Hubble Space Telescope supported
      the galaxy-galaxy interaction scenario by identifying that some
      of post-starburst (E+A) galaxies in high redshift clusters show
      disturbed or interacting signatures (Couch et al. 1994,1998;
      Dressler et al 1994; Oemler, Dressler, \& Butcher 1997).
           Liu \& Kennicutt (1995a,b) observed 40 merging/interacting
      systems and found that some of their spectra resemble E+A
      galaxies. Bekki, Shioya, \& Couch (2001) modeled galaxy-galaxy mergers with
    dust extinction, confirming that such systems can produce
    spectra which evolve into E+A spectra. 
  Recently, G03 found that young E+A galaxies have 8 times more
      companion galaxies within 50 kpc, providing a strong support for the
      merger/interaction origin of E+A galaxies.  

 However, an important issue remains to be addressed on  the origin of E+A galaxies. E+A galaxies may  be explained as dusty-starburst galaxies. In this scenario, E+A galaxies are not post-starburst galaxies, but in reality star-forming galaxies whose emission lines are invisible in optical wavelengths due to the heavy obscuration by dust.  
    As a variant,  Poggianti \& Wu (2000) presented the selective dust extinction
   hypothesis, where dust extinction is dependent on stellar age since 
    youngest stars inhabit very dusty star-forming HII regions while older
    stars have had time to migrate out of such dusty regions. 
     If O, B-type stars in E+A galaxies are embedded in dusty regions and
    only A-type stars have long enough lifetimes ($\sim$ 1 Gyr) to move out from such regions,  
    this scenario can naturally explain the E+A phenomena.
   A straightforward test for these scenarios is to observe in radio wavelengths
  where the dust obscuration is negligible.
  At 20cm radio wavelengths, the synchrotron radiation
  from electrons accelerated by supernovae can be observed. Therefore,
  in the absence of a radio-loud active nucleus, the radio flux of a
  star-forming galaxy can be used to estimate its current massive star
  formation rate (Condon 1992; Kennicutt 1998; Hopkins et al. 2003). 
    Smail et al. (1999) performed such a radio observation and found that
  among 8 galaxies detected in radio, 5 galaxies have strong Balmer
  absorption with no detection in [OII].
  They concluded that massive stars are currently forming in these 5 galaxies. 
    Owen et al. (1999) investigated the radio properties of galaxies in
  a rich cluster at $z\sim$0.25 (A2125) and found that optical line
  luminosities (e.g., H$\alpha$+[NII]) were often weaker than one would
  expect for the SFRs implied by the radio emission.
    Miller \& Owen (2001) observed radio continua of 15 E+A galaxies and
  detected moderate levels of star formation in only 2 of them. The star
  formation rates (SFRs) of these two galaxies, however, are 
  5.9 and 2.2 M$_{\odot}$ yr$^{-1}$, which are an order of magnitude
  smaller than those in Smail et al. (1999), consistent with normal to
  low SFR instead of starburst. 

 These results on these radio observation studies are somewhat irreconcilable, perhaps due
 to the contamination in the E+A samples. Previous samples of E+A galaxies
 were often selected based solely on [OII] emission and Balmer
 absorption lines either due to the high redshift of the samples or due to
 instrumental reasons. According to Goto et al. (2003d), such a selection of E+A
 galaxies without information on H$\alpha$ line would suffer from 52\%
 of contamination from H$\alpha$ emitting galaxies.

In this paper, we aim to investigate E+A galaxies with strong H$\delta$ absorption (E+A with H$\delta$ EW $>6\AA$ without [OII] or H$\alpha$ emission) are dusty-starburst or not.
Since our sample of E+A galaxies is selected
 from the Sloan Digital Sky Survey (SDSS; Abazajian et al. 2003) using both [OII] and H$\alpha$ lines,
 our sample is much larger and more complete than previous samples, without any contamination from H$\alpha$ emitting galaxies. In addition, we have selected E+A galaxies with stronger Balmer absorption lines than any previous samples.
 By observing this larger and better defined sample of E+A galaxies in the radio continuum at 20 cm using the Very Large Array (VLA), we aim to detect the dusty
 star formation in E+A galaxies.

 This paper is organized as follows: In
 Section \ref{data}, we describe the VLA observation;
 In Section \ref{Feb 15 09:43:03 2004}, we present the results;
  In Section \ref{discussion}, we discuss the physical implications of our results;
 In Section \ref{conclusion}, we summarize our work and findings.
   The cosmological parameters adopted throughout this paper are $H_0$=75 km
 s$^{-1}$ Mpc$^{-1}$, and
($\Omega_m$,$\Omega_{\Lambda}$,$\Omega_k$)=(0.3,0.7,0.0).

\section{VLA Observation}\label{data}

\subsection{Sample Selection}

\begin{figure*}
\begin{center}
\includegraphics[scale=0.25]{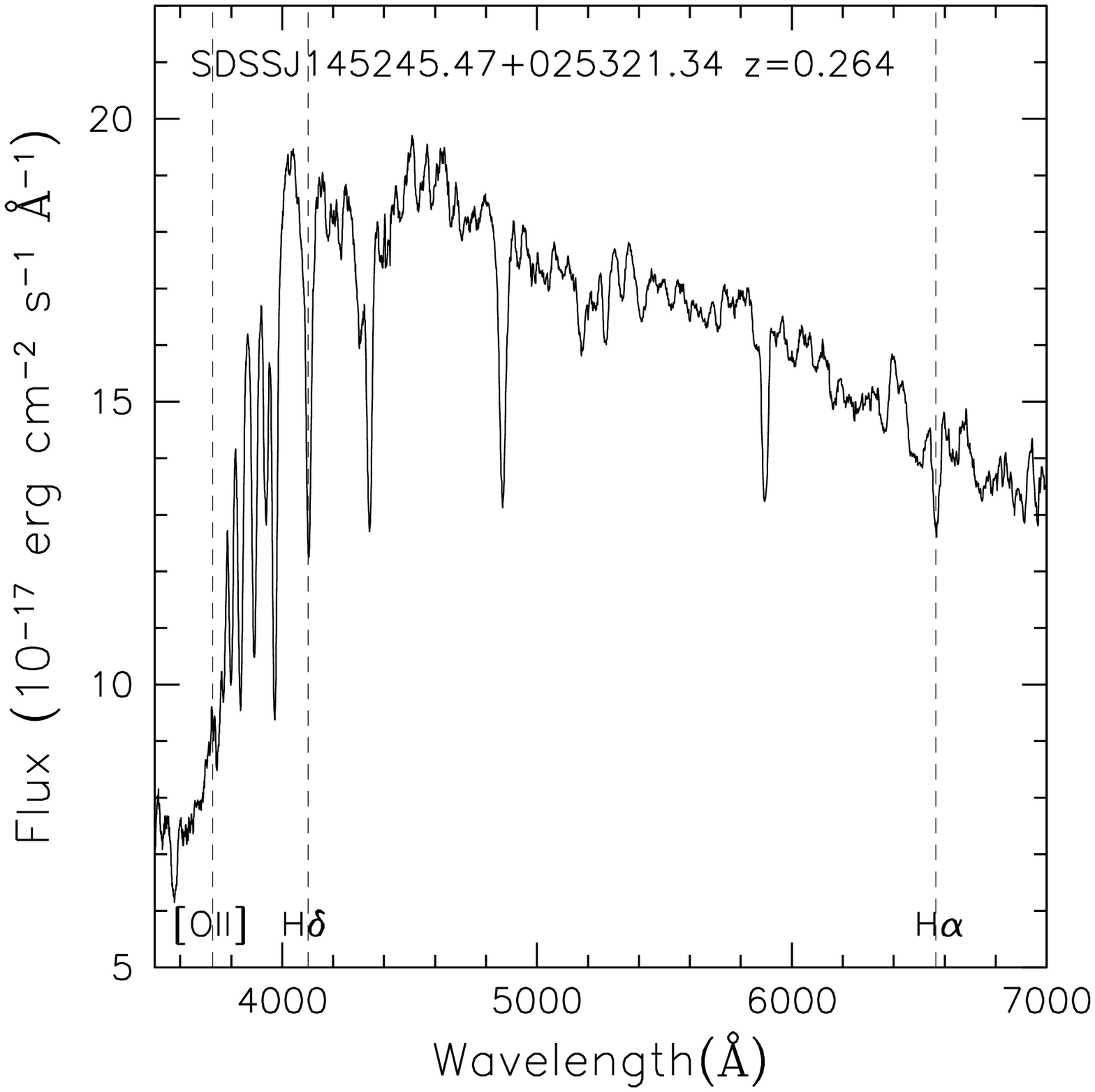}
\includegraphics[scale=0.25]{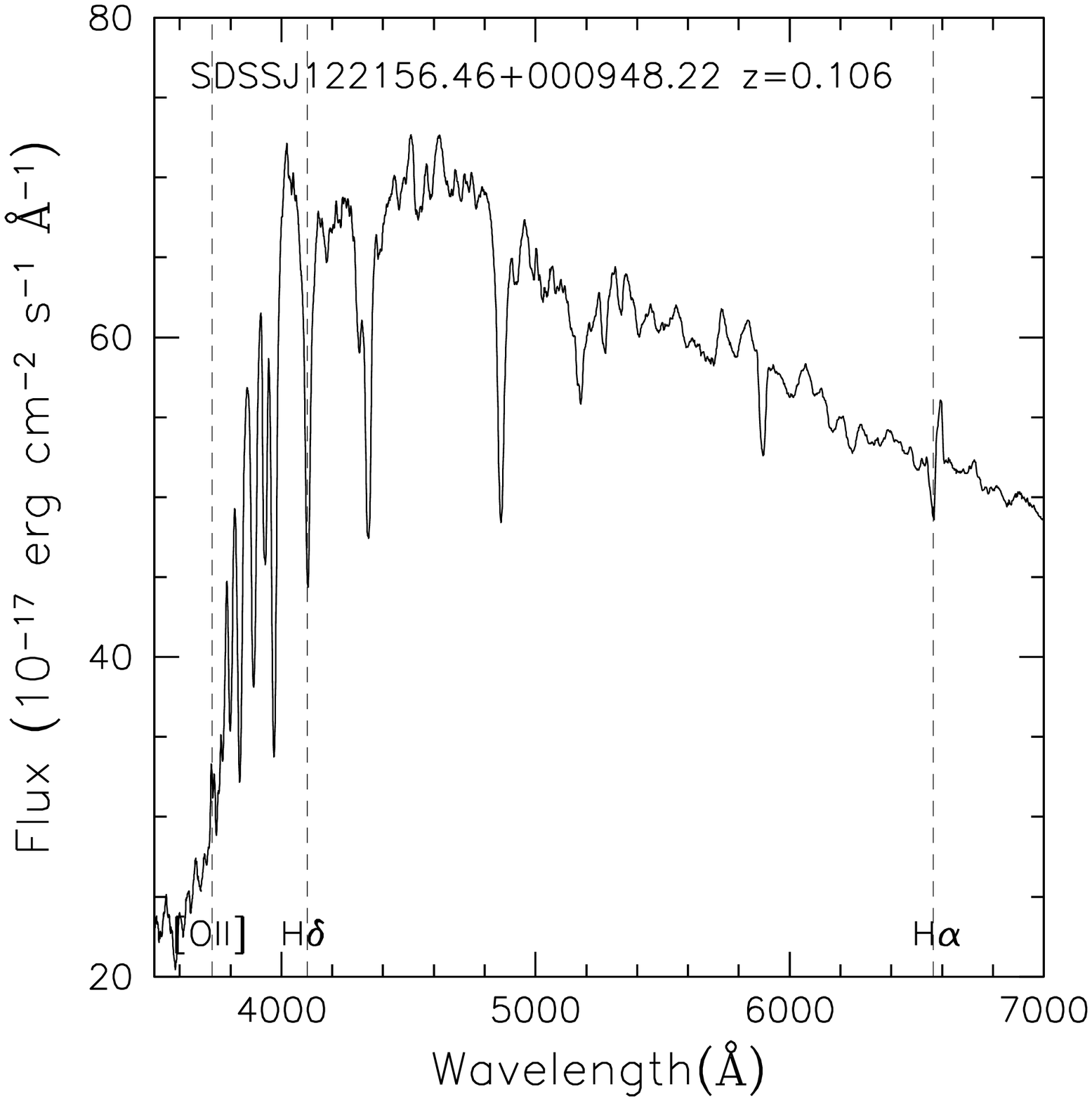}
\end{center}\begin{center}
\includegraphics[scale=0.25]{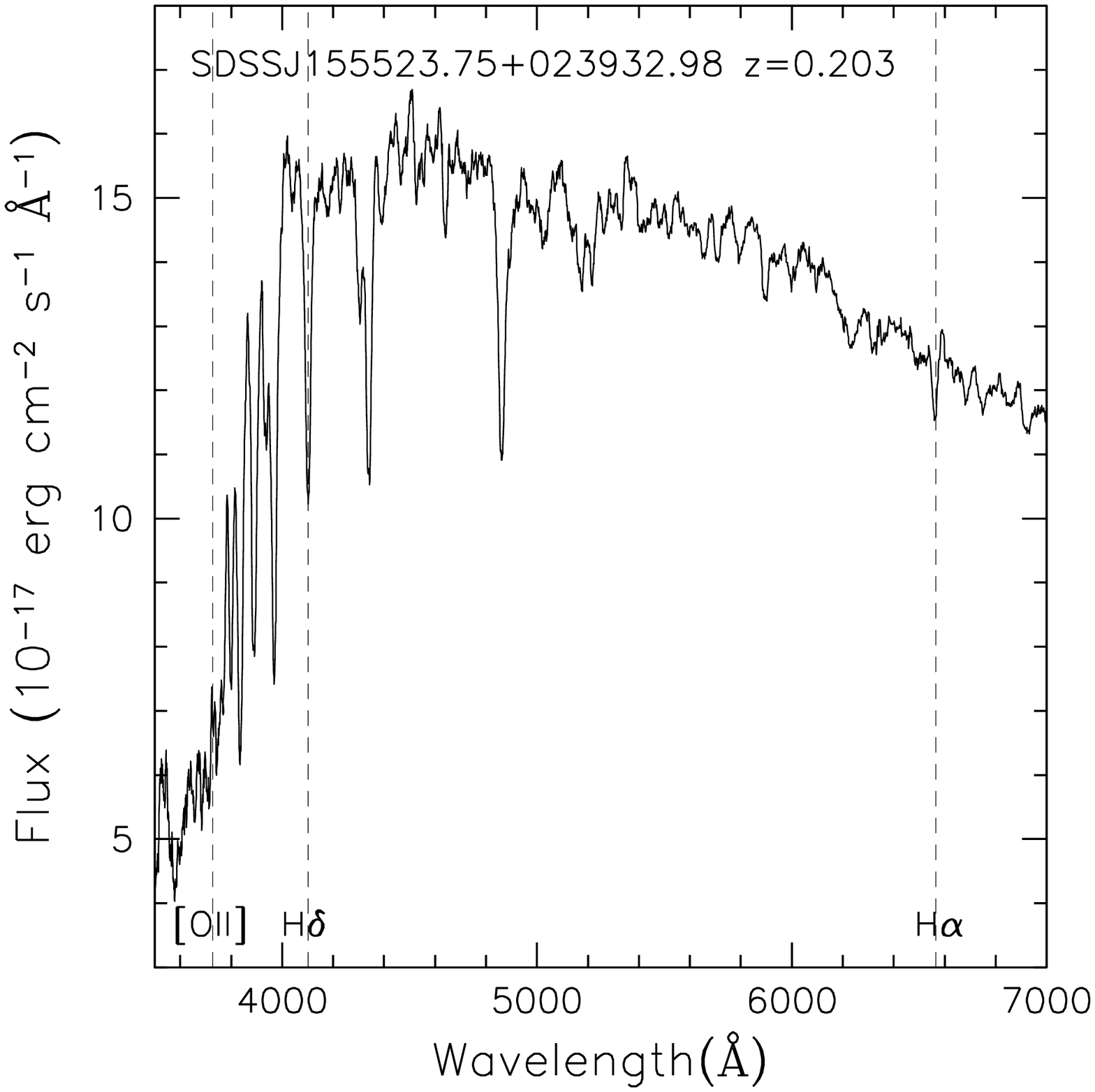}
\includegraphics[scale=0.25]{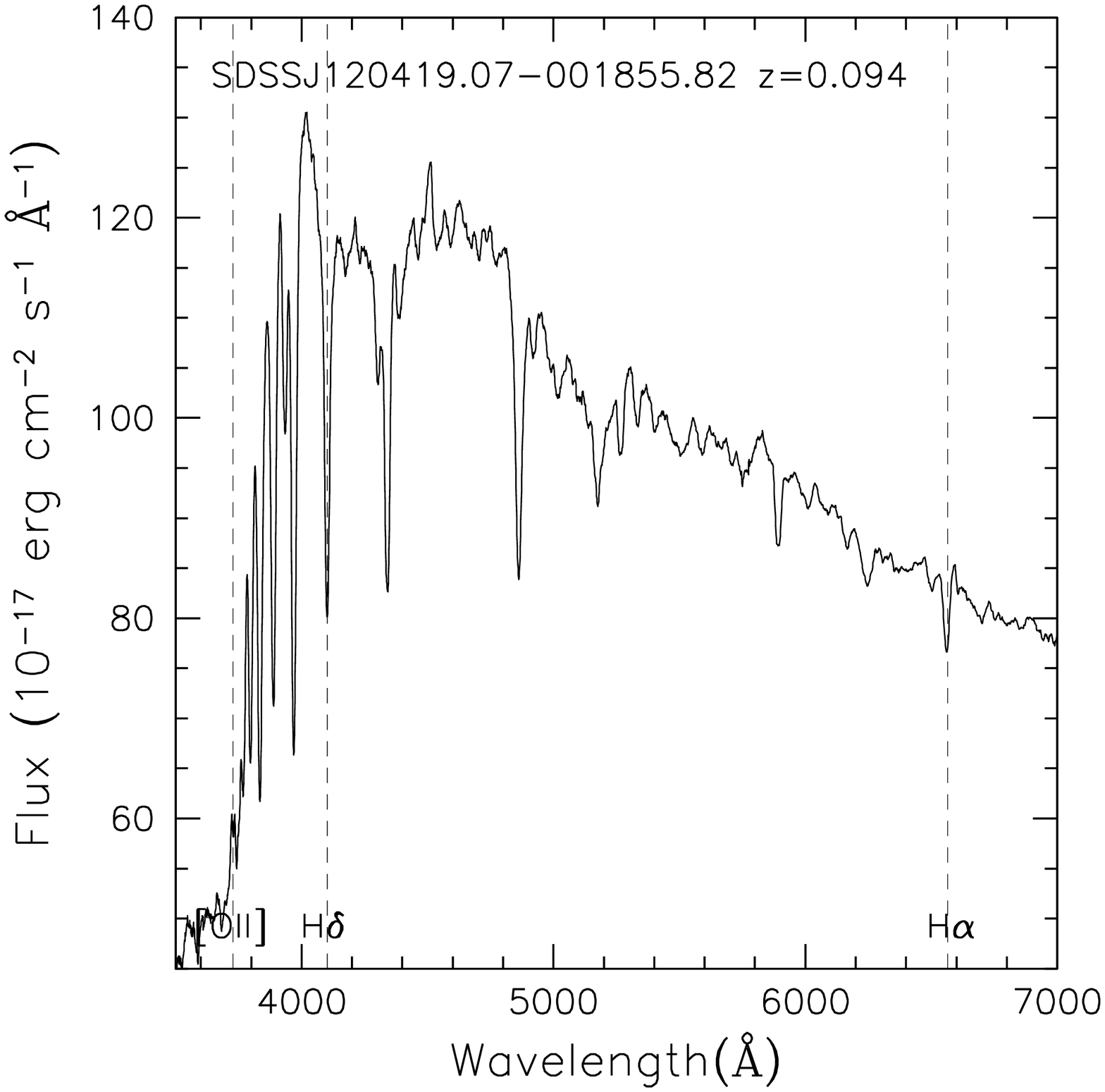}
\end{center}
\caption{Four example spectra of our target E+A galaxies (E+As with H$\delta$ EW $>$6 \AA). Spectra are shifted
 to restframe and smoothed using a 20\AA\ box. 
}\label{fig:ea2_spectra}
\end{figure*}

 We have selected our targets from the publicly available catalog of E+A
 galaxies described in Goto et al. (2003d). We have restricted our targets to those galaxies which have
 H$\delta$ equivalent width (in absorption) greater than 6 \AA\ and no
 detection of [OII] or H$\alpha$ emission lines within 1 $\sigma$.
 As an example, optical spectra of four of our sample E+A galaxies are shown in Fig. \ref{fig:ea2_spectra}. 
  Non-detection of [OII] or H$\alpha$ emission lines makes sure that these young E+A galaxies do not possess any current star formation at least based on optical information. The strong H$\delta$ absorption selects galaxies in post-starburst phase.
 Our targets include 23 E+A galaxies with H$\delta$ equivalent width (EW)
 $>$ 7\AA\, 
 and additional 13 E+A galaxies with  6 $<$ H$\delta$ EW $\leq$ 7\AA. The
 coordinates, redshift, H$\delta$ EW, and $M_g$ (absolute magnitude in $g$) of these targets are presented in Table 1. Since our targets are selected from the spectroscopic survey of the SDSS, galaxies at higher redshift tend to have brighter absolute magnitude.

\begin{figure}
\begin{center}
\includegraphics[scale=0.4]{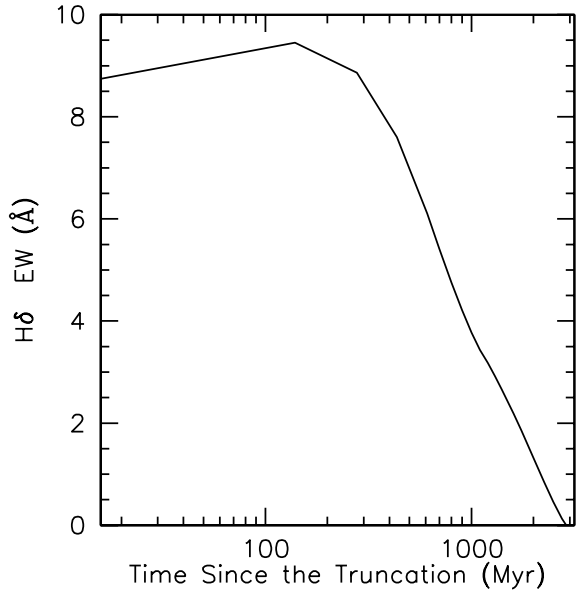}
\end{center}
\caption{
 H$\delta$ EWs are plotted against time (age) after the truncation for the 
 burst model (duration of 1 Gyr) of Bruzual \& Charlot (2003). 
 The model assumes Salpeter IMF and solar metallicity. }\label{fig:ea2_time_hd}
\end{figure}

 We would like to emphasize the advantage in selecting E+A galaxies with
 larger H$\delta$ EW. Compared with typical H$\delta$ EW $\sim$4\AA\ of
 previous E+A samples, our targets have much larger H$\delta$ EW of
 greater than 6\AA. By selecting E+A galaxies with larger H$\delta$ EW,
 we can select either younger E+A galaxies or E+A galaxies with stronger (with larger gas mass) previous
 starburst. 
If a larger fraction of a galaxy's total gas mass undergoes a starburst which is then truncated, the galaxy will have a stronger H$\delta$ absorption line.
 If
 an E+A galaxy is younger and closer to the truncation epoch, the galaxy
 has a stronger H$\delta$ absorption. In Fig. \ref{fig:ea2_time_hd}, we show H$\delta$ EW
 against time after the truncation using the burst model (duration of 1 Gyr before truncation) of
 Bruzual \& Charlot (2003) with Salpeter IMF and solar metallicity. In Fig. \ref{fig:ea2_time_hd}, the H$\delta$ EW becomes larger and larger after the galaxy truncates its star formation as O,B-type massive stars die.  
This leaves the HII regions
 without ionizing photons causing the Balmer emission features to
 disappear. 
After $\sim$200 Myr, A-type stars start to end their life and the H$\delta$ EW becomes close to zero after $\sim$1 Gyr.
 If we assume this  burst model (with 100\% burst
 strength, and a duration of 1 Gyr), the E+A galaxies with H$\delta$ EW$>$7 \AA\ have age of
 270-430 Myr after the truncation of the burst. 

 It is less understood how soon the 20cm radio emission ceases after the truncation of the star formation since the steps between star formation and synchrotron emission (e.g., super nova explosion, acceleration of relativistic electrons in the SNR, propagation of the cosmic rays throughout the galaxy, energy loss, and escape) are quite complicated. However, since the synchrotron emission is caused by the supernova explosion (Type II and Ib) of stars more massive than $\sim$8 M$_{\odot}$ (Condon 1992), it is expected that the 20cm emission does not continue beyond $\sim$100 Myr. It is estimated that the time lag between the peaks of the radio emission and other SFR indicators is less than 50 Myr (see Fig.~1 of Rengarajan \& Mayya 2003). Therefore, provided the absence of the dust-hidden star formation, our target E+A galaxies should not have significant radio emission in 20cm.

\subsection{Observation}

 We have observed these 36 E+A galaxies in 20cm continuum on January
 24,25,29,30, and February 10 in 2004 using the Very
 Large Array (VLA). The array was operated in the CnB configuration in the continuum mode.
 The exposure time was between 25 to 29 mins depending on the target. 
 For each target, a nearby
 phase calibrator was also observed twice for a few minutes. The primary flux
 calibrators (either 3C286 or 3C48 depending on the observing runs) which were adjusted to the flux scale of Baars 
 et al. (1977) were observed once for every observing run.
 Data reduction was performed with the AIPS using the standard steps. 
 Unfortunately, for most of the targets, the rms noise level was limited by the confusion from
 nearby sources, varying in the range of 0.1-1mJy.

\section{Results}
\label{Feb 15 09:43:03 2004}

None of our target E+A galaxies were detected at 20 cm except for the following two:  
(i) SDSSJ091227.78+534222.95 --- this target has a 17.5 mJy source at 5 arcsec away. The extended radio emission from the nearby source covers the target, making it impossible to estimate the radio flux at the target position; 
(ii) SDSSJ111108.08+004048.79 --- a 50.3 mJy source is located at 9.8 arcmin away from the target. The entire field of view had the high noise level due to this strong source. 
 We have excluded these two targets from the following analysis.

 The rest of the targets were not detected in our observation.
 The resulting rms noise levels at the target positions are presented in Table 1. Some targets are successfully observed to the low noise level ($<$0.1 mJy), however, other targets have high noise level ($\sim$1 mJy) due the confusion from nearby sources. 
Note that due to the different distance to the targets ($0.06<z<0.3$) and due to the different rms noise level of each exposure, the types of dusty-starburst galaxies that can be investigated are also different among the sample.
 Taking the 3 $\sigma$ of this noise level as an upper limit of the 20 cm flux, we have computed the upper limit of the radio luminosity of each target using the following equation (Petrosian \& Dickey 1973; Morrison et al. 2003 ApJS).
  \begin{equation} 
 L_{1.4GHz}=4 \pi D_L^2 S_{1.4GHz} (1+z)^{\alpha}/(1+z),
  \end{equation}
 where $D_L$ is a luminosity distance, $S_{1.4GHz}$ is the flux density, $(1+z)^{\alpha}$ is the color correction and  $1/(1+z)$ is the bandwidth correction. We assumed a SED shape of $S\propto \nu^{-\alpha}$ with $\alpha=0.8$,  which is appropriate
  for galaxies dominated by synchrotron emission (Condon 1992). The resulting upper limits of the radio powers are listed in Table 1. 

We have calculated the upper limit of the radio estimated SFR using the following conversion (Yun, Reddy \& Condon 2001).
  \begin{equation} 
   SFR (M_{\odot}\ yr^{-1})= 5.9 \times 10^{-22} L_{1.4 GHz} (W Hz^{-1}).
  \end{equation}
  This conversion assumes a Salpeter initial mass function integrated
  over all stars ranging from 0.1 to 100 M$_{\odot}$ and hence
  represents the total SFR of a galaxy. 
 Note that this conversion depends on the integrated range of the initial mass function (e.g., SFR integrated over 0.1-100 M$_{\odot}$ is 5.5 times larger than that over 5-100 M$_{\odot}$).
   The resulting upper limits on SFR are plotted against redshift in Fig. \ref{fig:ea2_radio_sfr}.
 Numerical values are provided in Table 1.
Except for  one E+A galaxy at $z\sim$0.3 for which we were unable to obtain a tight constraint on the SFR, 33 E+A galaxies have SFR lower than $\sim$ 100 $M_{\odot}\ yr^{-1}$. We have obtained tighter constraints for nearby E+A galaxies. Out of 19 E+A galaxies at $z<0.15$, 9 E+A galaxies have SFR lower than 10 $M_{\odot}\ yr^{-1}$. 

\section{Discussion}\label{discussion}

In Section \ref{Feb 15 09:43:03 2004}, none of
 our E+A sample had radio derived SFR $>100 M_{\odot}$ yr$^{-1}$ while 15 E+A's in
 our sample had no radio derived SFR $>15 M_{\odot}$ yr$^{-1}$.
Part of out result is apparently inconsistent with some previous work.  Miller \& Owen (2001) detected 2 galaxies out of 15 E+A galaxies. Smail et al. (1999) detected 5 galaxies out of 8 galaxies with  strong Balmer absorption with no detection in [OII]. However, Miller et al. (2001)'s sample was selected from the Las Campanas Redshift Survey (Shectman et al. 1996), whose spectrograph does not cover the H$\alpha$ line. Similarly, Smail et al. (1999)'s sample did not have information on the H$\alpha$ line due to the high redshift of their sample galaxies. Therefore, both of these E+A galaxies were selected only using the [OII] line as an indicator of the current SFR. According to Goto et al. (2003d), an E+A selection without information on the H$\alpha$ line can suffer from 52\% of contamination from H$\alpha$ emitting galaxies. Therefore, these E+A galaxies detected in 20 cm in previous work were probably not E+A galaxies, but the contamination from H$\alpha$ emitting galaxies.

\begin{figure}
\begin{center}
\includegraphics[scale=0.4]{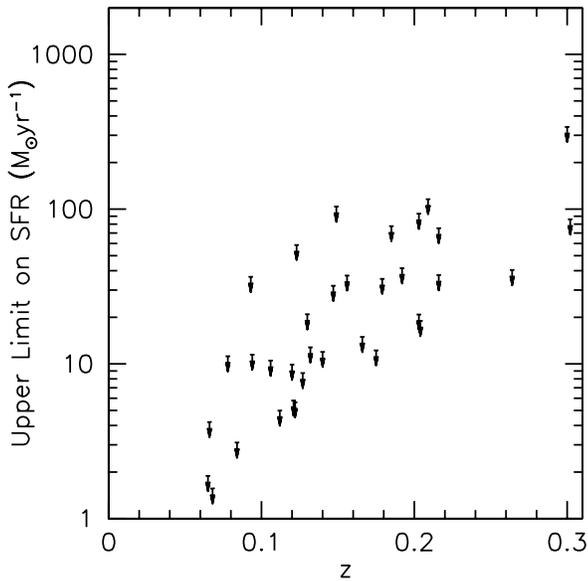}
\end{center}
\caption{ 
 Upper limits on SFR calculated using the radio 20cm continuum are plotted against redshift.
 None of our target galaxies were detected in the observation. Therefore, all data points show upper limit on the radio estimated star formation rate calculated from the 3 $\sigma$ of the rms sky noise. Note that the SFR is computed by integrating IMF over 0.1-100 M$_{\odot}$. The SFR over 5-100 M$_{\odot}$ is 5.5 times smaller than our value.
}\label{fig:ea2_radio_sfr}
\end{figure}

 More complete sample of E+A galaxies can lead us to the more robust conclusion. Our results of no radio detection in all 34 E+A galaxies support that E+A galaxies do not conceal a strong dusty starburst. Although we have not obtained useful constraints on 6 targets with log($L_{1.4GHz}$)$>$23, we have obtained the upper limit of the radio luminosity log($L_{1.4GHz}$)$<$22.3 for 15 targets (c.f.  the detection limit in Smail et al. 1999 and Owen et al. 1999 was log($L_{1.4GHz}$)$\sim$22.3).
For all the  E+A galaxies except three, we can exclude SFR $\geq$ 100 M$_{\odot}$ yr$^{-1}$ of dusty starburst (Fig. \ref{fig:ea2_radio_sfr}). For nearby sources at $z<0.08$, the upper limit of the SFR is even tighter at $\sim$ 10 M$_{\odot}$ yr$^{-1}$. 
The dusty starburst scenario where E+A galaxies are star-forming galaxies whose optical emission lines are invisible due to the dust obscuration, and the selective dust extinction scenario where O,B-type stars in E+A galaxies are embedded in dusty region, both looked like a plausible scenario. However, all of our results show evidence against these scenarios.At the very least, there are 7 E+A galaxies with the radio estimated SFR of $<$7 M$_{\odot}$ yr$^{-1}$. We cannot interpret these as strong starburst galaxies. 


 To this end, it is also important to  bear in mind that many E+A galaxies are found in the field region instead of cluster regions (e.g., Dressler et al. 1999) in recent large surveys of the nearby universe.  Although Dressler et al. (1999) claimed that the fraction of k+a/a+k galaxies is higher in clusters than in the field at $z=0.4$, 
 Quintero et al. (2003) found E+A galaxies having a similar local galaxy density as the field galaxies. Mateus \& Sodr{\'e} (2004) found that short-starburst galaxies are ubiquitous in the field regions. G03 also found that most of E+A galaxies are in the field regions. 
 An apparent inconsistency between Dressler et al. (1999) and the local results may be  due to their  sample selection without H$\alpha$ information, or different line criteria for the E+A selection, or the difference in the redshift. 
 However, it is clear that these field E+A galaxies found at low redshifts cannot be explained by cluster related physical mechanisms such as the ram-pressure stripping (e.g., Fujita \& Goto 2004) or tidal interaction (e.g., Gnedin 2003a,b). 

It has been long believed that E+A galaxies might be the transition object in the cluster galaxy evolution such as the Butcher-Oemler effect (e.g., Goto et al. 2003b) or the morphology-density relation (e.g., Goto et al. 2003c; 2004). However, many field E+A galaxies indicate that explaining cluster galaxy evolution through E+A galaxies is perhaps  not plausible any more. Passive spiral galaxies (Goto et al. 2003e; Yamauchi \& Goto 2004) may be an alternative candidate for the transition objects in cluster regions instead of E+As.

Since neither the dusty starburst or the cluster related mechanisms are likely to be the origin of E+A galaxies, we need to consider findings by G03 more seriously. They found that E+A galaxies have $\sim$8 times more companion galaxies within 50 kpc than normal galaxies do. Since their results are significant at 2 $\sigma$ level, they proposed that the origin of E+A galaxies are likely to be merger/interaction with close companion galaxies. Their finding of early-type morphology of E+A galaxies also supports this scenario since the merger/interaction remnants are known to have elliptical like morphology (e.g., Barnes \& Hernquist 1992). Based on the derived upper limits of radio estimated SFR in 34 E+A galaxies (Figure \ref{fig:ea2_radio_sfr}), the result of this work also points to the merger/interaction scenario.

\section{Conclusion}\label{conclusion}

In order to answer the question whether E+A galaxies are dusty starburst or not, we have performed a 20 cm radio continuum observation of 36 E+A galaxies with H$\delta$ EW $>6$\AA\ using the VLA. Except for the two sources with contamination from a nearby radio source, none of our 34 targets were detected in 20 cm radio continuum to the limit listed in Table 1. This result shows that our selection of E+A galaxies was successful in selecting galaxies in the post-starburst phase. We have calculated upper limits on the radio estimated SFR using the 3 $\sigma$ of the rms noise level as shown in Fig. \ref{fig:ea2_radio_sfr}.
At  $z<0.3$, all the E+A galaxies observed have an upper limit of the SFR less than $\sim100 M_{\odot}$ yr$^{-1}$. Out of 19 E+A galaxies at $z<0.15$, 9 E+A galaxies have SFR lower than 10 $M_{\odot}\ yr^{-1}$. 
These results indicate that E+A galaxies are not likely to possess an on-going strong starburst hidden by the dust extinction.

\section*{Acknowledgments}

We thank the anonymous referee for many insightful comments, which have improved the paper significantly.
We are indebted to Mark Claussen for his friendly help during the VLA
observation and the data reduction.
 The VLA is operated by the National Radio Astronomy Observatory, which
 is a facility of the National Science Foundation operated under
 cooperative agreement by Associated Universities, Inc.

\clearpage

\clearpage

\begin{table*}
\begin{center}\label{tab:radio_sfr} 
\centerline{\sc \hfil Table 1\hfil }
\centerline{\sc \hfil List of Targets\hfil }
{\small\scriptsize
\hspace{-1.truein}\begin{tabular}{lccrcccrrr}
\noalign{\medskip}
\hline\hline
\noalign{\smallskip}
{Name} & {R.A.} & {Dec.}  & {$z$} &H$\delta EW $ &  $M_g$  & rms noise   &  Upper limit of log $P_{1.4GHz}$  & SFR upper limit $^{\ast}$ &  \hfil  \cr
 &  &   &  & (\AA) &   &  (mJy)  &  (log W Hz$^{-1}$) & (M$_{\odot}$ yr$^{-1}$ )   \hfil  \cr
\noalign{\smallskip}
\noalign{\hrule}
\noalign{\smallskip}
SDSSJ012250.33+151546.22  & 01:22:50.33 & +15:15:46.22 &    0.132 &   9.35 & -19.94 & 0.14 &  22.33 & 12.7       \\ 
SDSSJ014447.13+003214.89  & 01:44:47.13 & +00:32:14.89 &    0.179 &   7.79 & -22.16 &0.19  &  22.77 & 35.4       \\ 
SDSSJ030428.04$-$084906.62  & 03:04:28.04 & $-$08:49:06.62 &    0.120 &   7.03 & -21.01 & 0.13 &  22.22 & 9.8    \\ 
SDSSJ031911.86$-$003028.01  & 03:19:11.86 & $-$00:30:28.01 &    0.147 &   7.38 &  -20.50 & 0.27 &  22.73 & 32.0  \\ 
SDSSJ032551.21+003114.54  & 03:25:51.21 & +00:31:14.54 &    0.149 &   7.50 & -20.66 & 0.87 &  23.24 &103.9  \\ 
SDSSJ090424.90+535614.07  & 09:04:24.90 & +53:56:14.07 &    0.140 &   6.92 & -22.87 & 0.11   &22.30 & 11.9  \\ 
SDSSJ091227.78+534222.95  & 09:12:27.78 & +53:42:22.95 &    0.222 &   7.87 &  -22.55 & 1.87 & $\cdots$&$\dagger$ \\ 
SDSSJ092251.59+583746.75  & 09:22:51.59 & +58:37:46.75 &    0.166 &   6.69 & -21.48 &   0.09  & 22.40 & 14.9  \\ 
SDSSJ092344.28+575203.21  & 09:23:44.28 & +57:52:03.21 &    0.156 &   7.16 &-20.59& 0.28 &  22.80 & 37.2 \\ 
SDSSJ092509.94+042907.12  & 09:25:09.94 & +04:29:07.12 &    0.185 &   7.67 &-21.45& 0.38 &  23.11 & 77.7 \\ 
SDSSJ094347.55+020557.95  & 09:43:47.55 & +02:05:57.95 &    0.093 &   8.36 &  -22.31&0.92 &   22.79 &36.4  \\ 
SDSSJ094701.49+600720.13  & 09:47:01.49 & +60:07:20.13 &    0.123 &   7.85 & -20.84&0.77 &  22.99& 58.7 \\ 
SDSSJ101519.69+010341.63  & 10:15:19.69 & +01:03:41.63 &    0.216 &   7.32 & -21.58 & 0.25 &    23.10 & 75.4 \\ 
SDSSJ111108.08+004048.79  & 11:11:08.08 & +00:40:48.79 &    0.184 &   7.33 & -22.46& 3.39 & $\cdots$& $\ddagger$ \\ 
SDSSJ115037.52+013000.57  & 11:50:37.52 & +01:30:00.57 &    0.078 &   7.11 & -19.80&0.42 &  22.27 & 11.1 \\ 
SDSSJ120419.07$-$001855.82  & 12:04:19.07 & $-$00:18:55.82 &0.094 &   8.07 & -20.25&0.28 &  22.28& 11.4 \\ 
SDSSJ122156.46+000948.22  & 12:21:56.46 & +00:09:48.22 &    0.106 &   8.19 & -22.81 & 0.19 &  22.25 & 10.4 \\ 
SDSSJ124252.96+023700.90  & 12:42:52.96 & +02:37:00.90 &    0.084 &   6.52 &-20.83& 0.09 &  21.71 & 3.0  \\ 
SDSSJ132315.58+630726.73  & 13:23:15.58 & +63:07:26.73 &    0.175 &   6.86 & -22.35& 0.06 & 22.31 & 12.2 \\ 
SDSSJ133757.98+654410.48  & 13:37:57.98 & +65:44:10.48 &    0.066 &   7.82 & -22.09&0.23 &   21.85 & 4.2\\ 
SDSSJ134802.18+020405.73  & 13:48:02.18 & +02:04:05.73 &    0.068 &   6.41 & -21.78&0.08 &   21.42 & 1.5 \\ 
SDSSJ143137.97+021942.27  & 14:31:37.97 & +02:19:42.27 &    0.065 &   6.38 & -20.79& 0.10 & 21.50& 1.8  \\ 
SDSSJ143414.02+030140.28  & 14:34:14.02 & +03:01:40.28 &    0.302 &   7.48 & -20.46& 0.11  &  23.16 & 85.9  \\ 
SDSSJ143449.18$-$000919.95  & 14:34:49.18 & $-$00:09:19.95 &0.130 &   7.21 & -20.64 & 0.24 & 22.55 & 20.9 \\ 
SDSSJ143945.23+013501.79  & 14:39:45.23 & +01:35:01.79 &    0.121 &   6.49 & -21.64& 0.07  &   21.99 & 5.7\\ 
SDSSJ145245.47+025321.34  & 14:52:45.47 & +02:53:21.34 &    0.264 &   8.58 & -22.50& 0.07 &  22.83 & 40.4 \\ 
SDSSJ152847.79+032321.28  & 15:28:47.79 & +03:23:21.28 &    0.300 &   7.03 & -20.13&0.47 & 23.76 & 340.7 \\ 
SDSSJ155243.17+544847.50  & 15:52:43.17 & +54:48:47.50 &    0.204 &   6.16 & -21.23 & 0.07  &  22.50 & 18.9 \\ 
SDSSJ155523.75+023932.98  & 15:55:23.75 & +02:39:32.98 &    0.203 &   8.08 & -20.67 & 0.36 &  23.19 &93.3 \\ 
SDSSJ161016.75+484327.85  & 16:10:16.75 & +48:43:27.85 &    0.203 &   6.50 & -21.74& 0.08 &  22.54 &20.8 \\ 
SDSSJ163835.05+440349.44  & 16:38:35.05 & +44:03:49.44 &    0.112 &   7.45 & -21.92 & 0.08 &   21.92 &4.9 \\ 
SDSSJ171040.62+572348.28  & 17:10:40.62 & +57:23:48.28 &    0.122 &   6.08 & -20.87 & 0.07 &   21.98& 5.6\\ 
SDSSJ171546.26+582255.47  & 17:15:46.26 & +58:22:55.47 &    0.127 &   7.37 & -20.14 & 0.10 &  22.16& 8.7\\ 
SDSSJ215459.13$-$065114.40  & 21:54:59.13 & $-$06:51:14.40 &    0.216 &   7.62 & -21.00 & 0.12 & 22.80& 37.4 \\ 
SDSSJ225656.76+130402.59  & 22:56:56.76 & +13:04:02.59 &    0.209 &   8.13 &-21.25&  0.42 &  23.29& 116.1 \\ 
SDSSJ234711.33+005653.17  & 23:47:11.33 & +00:56:53.17 &    0.192 &   7.28 & -20.93 &0.18 &  22.84 &41.5 \\ 
\hline
\end{tabular}
}
\end{center}
\begin{flushleft}
$\ast$ --The SFR is computed by integrating IMF over 0.1-100 M$_{\odot}$, which returns 5.5 times larger SFR than that in some previous work (Smail et al. 1999; Owen et al. 1999) where the integration is over 5-100 M$_{\odot}$.\\
$\dagger$ -- An extended emission from a 17.5 mJy source at 5 arcsec away covers the target.\\
$\ddagger$ -- A 50.3 mJy source is located at 9.8 arcmin away.\\
\end{flushleft}
\end{table*}

\end{document}